\documentclass[10pt,journal]{IEEEtran}
\usepackage[pdftex]{graphicx}
\usepackage{algorithm}
\usepackage{color}
\usepackage{caption}
\usepackage{algpseudocode}
\newcommand{\RNum}[1]{\uppercase\expandafter{\romannumeral #1\relax}}
\definecolor{ForestGreen}{RGB}{162,52,0}
\usepackage{xcolor}
\usepackage{amsmath}
\usepackage{CJK}
\usepackage{url}
\usepackage{array}
\usepackage{ragged2e}
\usepackage{latexsym}
\usepackage{ifsym}

\ifCLASSOPTIONcompsoc
  \usepackage[nocompress]{cite}
\else
  \usepackage{cite}
\fi
\ifCLASSINFOpdf
\else
\fi
\ifCLASSOPTIONcompsoc
  \usepackage[caption=false,font=footnotesize,labelfont=sf,textfont=sf]{subfig}
\else
  \usepackage[caption=false,font=footnotesize]{subfig}
\fi

\begin{document}

\title{Resource Management and Security Scheme of ICPSs and IoT Based on VNE Algorithm}

\author{Peiying Zhang, Chao Wang, Chunxiao Jiang, Neeraj Kumar, and Qinghua Lu
\thanks{This work is partially supported by the National Key Research and Development Program of China under Grant 2020YFB1804800, partially supported by the National Natural Science Foundation of China under Grant 61922050, and partially supported by Shandong Provincial Natural Science Foundation under Grant ZR2020MF006. \textit{(Corresponding author: Chunxiao Jiang).}}
	\thanks{P. Zhang and C. Wang are with the College of Computer Science and Technology, China University of Petroleum (East China), Qingdao 266580, China. E-mail: 25640521@qq.com, wangch\_upc@qq.com.}
    \thanks{Chunxiao Jiang is with the School of Information Science and Technology, Tsinghua University, Beijing 100084, China. E-mail: jchx@tsinghua.edu.cn.}
	\thanks{N. Kumar is with the Department of Computer Science and Engineering, Thapar Institute of Engineering and Technology. Email: neeraj.kumar@thapar.edu.}
    \thanks{Qinghua Lu is with Data61, CSIRO, Sydney, Australia. E-mail: qinghua.lu@data61.csiro.au.}
}

\markboth{IEEE Internet of Things Journal}%
{Shell \MakeLowercase{\textit{et al.}}: Bare Demo of IEEEtran.cls for IEEE Journals}

\maketitle	

\begin{abstract}
The development of Intelligent Cyber-Physical Systems (ICPSs) in virtual network environment is facing severe challenges. On the one hand, the Internet of things (IoT) based on ICPSs construction needs a large amount of reasonable network resources support. On the other hand, ICPSs are facing severe network security problems. The integration of ICPSs and network virtualization (NV) can provide more efficient network resource support and security guarantees for IoT users. Based on the above two problems faced by ICPSs, we propose a virtual network embedded (VNE) algorithm with computing, storage resources and security constraints to ensure the rationality and security of resource allocation in ICPSs. In particular, we use reinforcement learning (RL) method as a means to improve algorithm performance. We extract the important attribute characteristics of underlying network as the training environment of RL agent. Agent can derive the optimal node embedding strategy through training, so as to meet the requirements of ICPSs for resource management and security. The embedding of virtual links is based on the breadth first search (BFS) strategy. Therefore, this is a comprehensive two-stage RL-VNE algorithm considering the constraints of computing, storage and security three-dimensional resources. Finally, we design a large number of simulation experiments from the perspective of typical indicators of VNE algorithms. The experimental results effectively illustrate the effectiveness of the algorithm in the application of ICPSs.
\end{abstract}


\section{Introduction}\label{part1}

\begin{figure}[!htp]
\centering
\includegraphics[width=1\columnwidth]{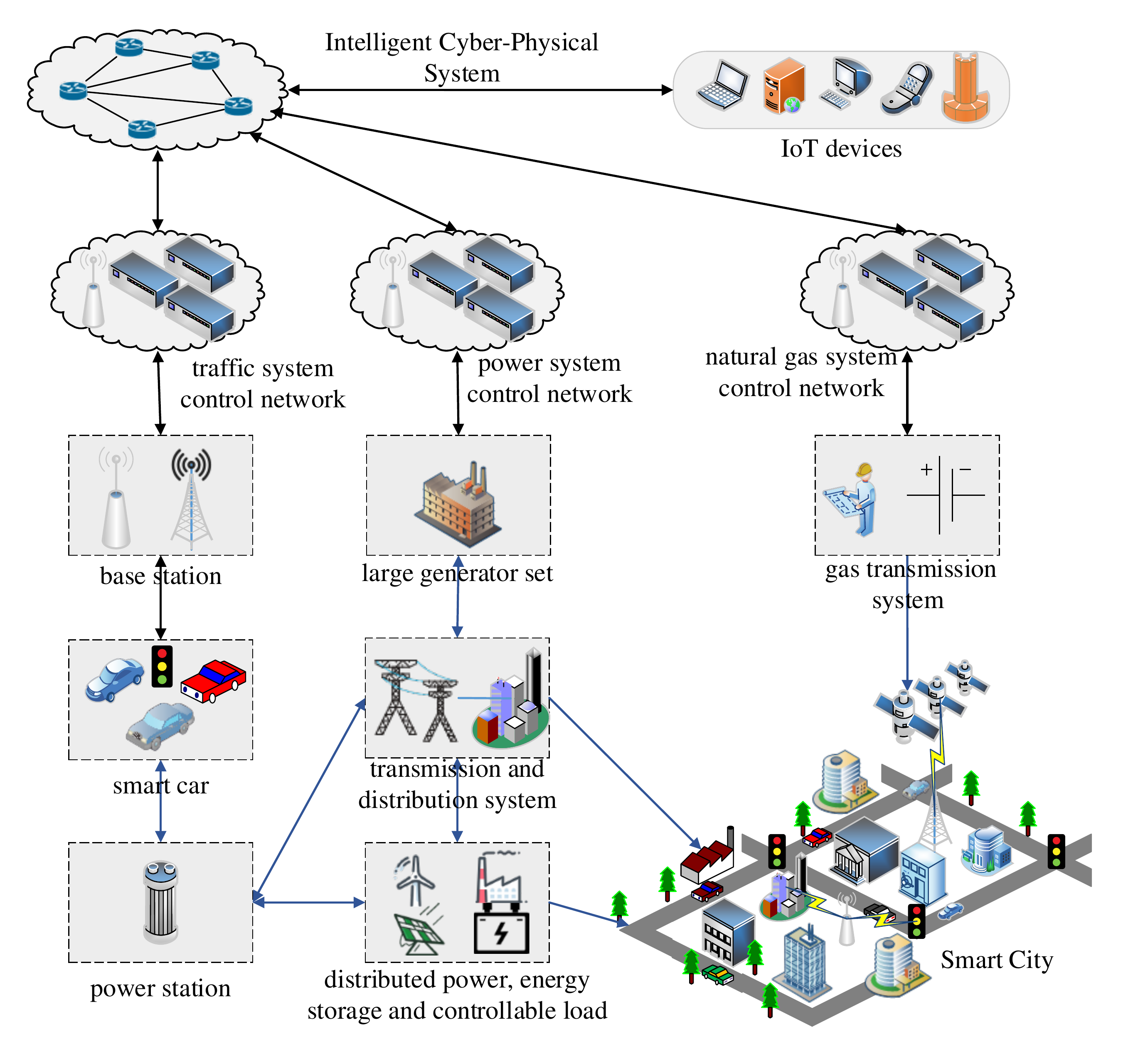}
\caption{Intelligent Cyber-Physical systems and IoT scenarios.}
\label{fig_ICPS}
\end{figure}

The emergence of Intelligent Cyber-Physical Systems (ICPSs) has completely subverted the way humans interact with intelligent systems \cite{c5,c6}. An ICPS is essentially a computer system, which uses various intelligent technologies to closely combine physics and software to achieve diversified operation modes at different time and space scales \cite{n1,c1,d6}. Because ICPSs have excellent characteristics of extensive use of intelligent technologies (machine learning (ML), reinforcement learning (RL), etc.), they are more valued in the field of Internet of Things (IoT). For example, ICPSs are applied in IoT scenarios such as intelligence transportation, smart home, smart grid and smart medical care. The scene of the ICPS is shown in Figure \ref{fig_ICPS}. What needs researchers to pay attention to is that ICPSs are facing severe challenges in data storage, resource optimization and network security \cite{d10}. In order for the ICPSs to match the development speed of new technologies and further improve the performance of supporting the construction of IoT, a new solution needs to be considered from the perspective of the underlying network architecture. Among them, resource optimization and scheduling are the most important issues for the underlying network architecture \cite{j4,d7,n2,d8}. The reasonable allocation of underlying network resources to ICPSs will help solve the new challenges faced by the development of the system \cite{j3,c2}. Radio network resource management faces severe challenges, including storage, spectrum, computing resource allocation, and joint allocation of multiple resources \cite{jcx1,jcx2}. With the rapid development of communication networks, the integrated space-ground network has also become a key research object \cite{jcx3}.

The large-scale application of IoT has profoundly changed people's lifestyles \cite{c7,c8}. The urgent performance requirements of massive IoT end users for IoT force it to require extremely robust infrastructure support. As the traditional Internet architecture continues to be rigid with the expansion of user scale, and the continuous development of artificial intelligence technologies bring great pressure to the resource allocation of the traditional Internet architecture \cite{c9,c10}. Therefore, its "best effort" service model cannot adapt to the development needs of the IoT, and it cannot be used as the underlying architecture to support the development of multi-dimensional applications of the IoT. The emergence of network virtualization (NV) provides new possible ideas for solving the basic network architecture problems that support the development of ICPSs \cite{z1}. In NV architecture, the realization of various network functions no longer depends on specific hardware but on software programming \cite{n3}. A key problem is how the infrastructure provider (InP) reasonably allocates the substrate node and link resources for the service provider (SP), i.e. the virtual network embedding (VNE) problem.

In ICPSs, host-centric networks are gradually changing to resource-centric networks. Therefore, resource scheduling capabilities are the core issue of ICPSs. Many challenges faced by ICPSs can be solved from the perspective of resource scheduling \cite{d9}. NV technology can improve the resource management capabilities of the network to a large extent, so it can be considered for resource management and security issues of ICPSs. As one of the key problems of NV, the efficiency of VNE algorithm largely determines the performance of ICPSs \cite{c12}. ICPSs need to handle complex tasks such as heterogeneous data generation, massive data transmission, fast changing cache state, etc. VNE algorithm can provide a good technical support for ICPSs to deal with these tasks \cite{b1,j1}.

With the rise of network intelligent learning algorithms, using intelligent algorithms to solve high-latitude decision-making problems has become a trend. It is precisely because of the excellent performance of RL that we apply it to solve the resource management problems of ICPSs and IoT \cite{c13,c14,c15}. In RL, agent takes new actions to get more rewards through interaction with the environment, which is a process of seeking the best \cite{c4,a2,n6}. In this paper, we propose a VNE algorithm based on RL method for computing, storage and security of three dimensional resource constraints. To this end, the main contributions of this paper are as follows:
\begin{enumerate}
\item Aiming at the resource management and security problems faced by ICPSs, this paper proposes a VNE algorithm constrained by CPU, storage resources and security attributes, which solves the challenges faced by ICPSs from the perspective of network resource allocation.

\item We use RL to improve the performance and efficiency of the algorithm. In RL method, we use a policy network as an agent to participate in training. The goal of agent is to explore the optimal VNE strategy. In test phase, the VNE process is carried out according to the training results.

\item In the experimental stage, we compared the performance of the algorithm proposed in this paper with several other representative algorithms. The experimental results prove that the algorithm proposed in this paper has good characteristics.
\end{enumerate}

The other chapters of the paper are arranged as follows. In section \ref{part2}, after a brief analysis of the problems to be studied, the related research progress of VNE algorithms based on network resource constraints is introduced. Section \ref{part3} describes the basic VNE problem, and then we create the network model for this problem. Finally, we mention the evaluation index of the VNE algorithm in this section. In section \ref{part4}, we introduce in detail the implementation process of the VNE algorithm based on RL for three-dimensional resource constraints. Section \ref{part5} introduces the details of the simulation experiment, and demonstrated and analyzed the experimental results. Section \ref{part6} summarizes the full paper.

\section{Related Work}\label{part2}

ICPSs have received widespread attention in the industry, and a considerable number of scholars have conducted representative studies. Zhou et al. \cite{d1} proposed a new ICPS based on IoT. The authors also studied the resource allocation of the system and innovatively introduced the concept of virtual devices. Through outage analysis, the authors transformed the mixed probability problem into a non-probability problem, and the final result proved that the method has good convergence characteristics. Reference \cite{d2} proposed a collaborative control method of network resources in ICPSs. This method mainly achieved system balance by controlling the performance of each loop, and ultimately optimized the overall performance of ICPSs. In addition, references \cite{d3} and \cite{d4} also studied the application of ICPSs in IoT.

In issues related to network resource allocation, CPU resource is one of the most important resources to realize network function. Besides, with the blowout growth of ICPSs, the security problems in the VNE are increasingly apparent \cite{a3,n4}. However, there is no special security protection mechanism to ensure the security of ICPSs under the NV architecture \cite{n5}. In addition, the existing virtual network model allows to exchange bandwidth resources with storage resources to a certain extent \cite{3,j2}, so it increases the difficulty of rational allocation of resources. Therefore, reasonable allocation of computing resources, storage resources and full consideration of the security of VNE are of great significance to solve the problem of VNE.

At present, some scholars have discussed the above-mentioned resource-constrained VNE algorithms, but most of them have designed and implemented heuristic algorithms from the perspective of a single resource. In addition, more and more VNE algorithms based on intelligent learning algorithms have appeared. This part will introduce some representative algorithms.

\subsection{Heuristic VNE Algorithm With Resource Constraints}

According to the concept of security virtual network embedding (SVNE), a new solution of SVNE was proposed by Liu et al. \cite{1}. This algorithm was implemented based on multi-attribute evaluation and path optimization. The node mapping function was used to map the virtual nodes by using the multiple characteristics of resource richness and security attributes. After that, it considered the characteristics of link bandwidth and path hops and used link mapping function to complete the process of link mapping. In reference \cite{2}, trust relationship and trust degree were introduced into the VNE problem. The virtual nodes were embedded by the method of approaching ideal ordering and the links were mapped by the shortest path method. In reference \cite{3}, the authors proposed a VNE algorithm based on storage, network constraints and computing resources. For the first time, the algorithm considered the influence of the above three important network resource constraints on the VNE algorithm. Two heuristic algorithms based on greedy node mapping were designed. Our algorithm is different from this one. First of all, we use RL algorithm to solve the resource constraint in VNE. In addition, we consider the problem of computing, storage and security resource constraints. This algorithm does not consider security issues.

\subsection{Embedded Algorithm of Virtual Network Based on Intelligent Learning}

The authors of \cite{4} tried to reduce VNE as a combinatorial optimization problem. The authors used the pointer network to establish a VNE strategy and used an active search algorithm to optimize the strategy. Experiments showed it can improve the utilization of network resources while increasing the long term revenue-cost ratio of VNE. In reference \cite{5}, the authors proposed a continuous decision-making VNE algorithm based on RL. In this algorithm, the VNR sequence was modeled as a seq2seq time series model, with emphasis on the continuous characteristics of node embedding. The experimental results proved that the algorithm has good VNE characteristics whether it is compared with heuristic algorithms or ML algorithm. Yuan et al. \cite{6} carried out research on VNE issues in the field of cloud computing. Aiming at the problems that most of the existing algorithms used random network topology and low resource utilization, the authors designed a VNE algorithm based on Q-learning. The characteristic of the algorithm was that the reward function was designed by the RL agent according to the embedding effect of the virtual links. The Q-matrix was updated through intelligent learning method, and then the agent can find the optimal embedding strategy from the latest Q-table. The authors of reference \cite{d5} proposed a new type of automatic VNE solution. In this scheme, the authors used a parallel deep reinforcement learning (DRL) framework and a multi-objective reward function method, and combined DRL with a neural network to achieve the goal of providing automatic embedding scheme within an acceptable time range.

Similarly, references \cite{8,9} also had introduced some ideas of ML algorithms into the VNE algorithm. All the above studies have achieved certain experimental results. But our work is different from these studies. First of all, we mainly take the network resource management and network security of ICPSs as the research background, and design an algorithm to solve these problems from the perspective of VNE. Secondly, we focus on the constraints of computing, storage and security of three-dimensional resources in VNE, and use RL method as a means to improve the efficiency of the algorithm. In the existing research on ICPSs resource management, VNE algorithm is not fully used. The existing VNE algorithm does not focus on computing, storage and security resource constraints. Therefore, our work is innovative.

\section{VNE Related Problem Description}\label{part3}

\subsection{Network Model}

Use graph theory to build mathematical models for substrate network and virtual network. The substrate network is represented by $G^S=\{N^S,L^S,A_N^S,A_L^S\}$. $N^S$ refers to all substrate nodes and $L^S$ refers to all substrate links. $A_N^S$ represents the collection of properties of the substrate node, including computing resource $CPU(n^s)$, storage resource $STO(n^s)$ and security level $SL(n^s)$, where $n^s \in N^S$. $A_L^S$ represents the substrate link attribute, including bandwidth resource $BW(l^s)$, where $l^s \in L^S$.

In the same way, the virtual network is represented by $G^V=\{N^V,L^V,A_N^V,A_L^V\}$. $N^V$ refers to all substrate nodes and $L^V$ refers to all substrate links. $A_N^V$ represents the collection of properties of the virtual node, including computing resource requirement $CPU(n^v)$, storage resource requirement $STO(n^v)$ and security level requirement $SR(n^v)$, where $n^v \in N^V$. $A_L^V$ represents the virtual link attribute, including the bandwidth resource requirement $BW(l^v)$, where $l^v \in L^V$. Therefore, the VNE problem can be expressed as $G_i^V \to G^S $, where $G_i^V$ represents one of the VNRs.

\subsection{VNE Problem Description}

\begin{figure}[!htp]
\centering
\includegraphics[width=1\columnwidth]{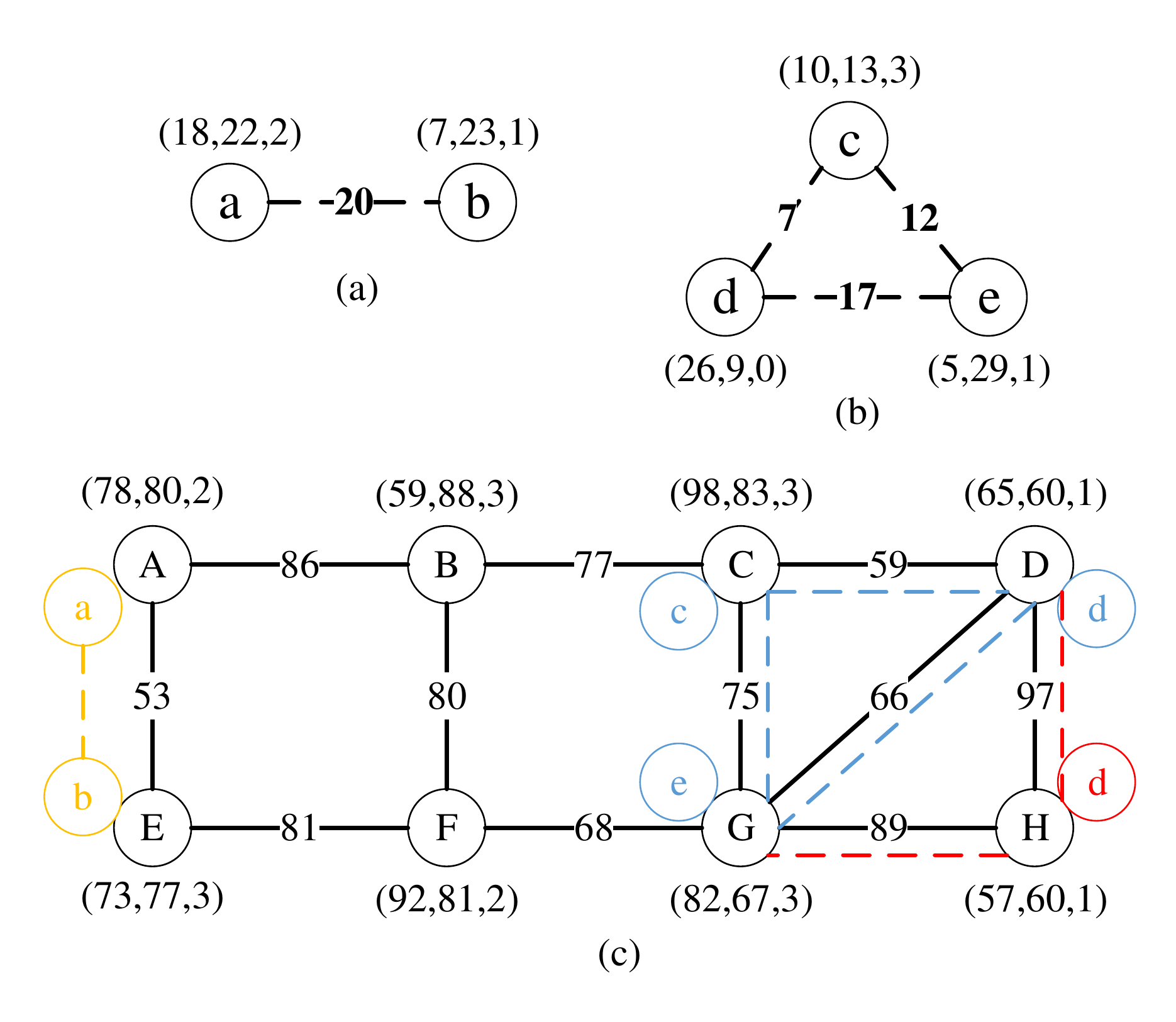}
\caption{Substrate network model, virtual network model and possible embedding examples.}
\label{fig_1}
\end{figure}

In Figure \ref{fig_1}, we show the schematic diagram of two VNRs (virtual network \textit{(a)} and virtual network \textit{(b)}) and a substrate network, which also shows the different embedding situations of VNRs. For two VNRs, the numbers in brackets beside the virtual nodes represent the computing resource requirements, storage resource requirements and security level requirements. The rest numbers on the virtual links represent the bandwidth requirements of the virtual links. In substrate network \textit{(c)}, the numbers in brackets represent the available computing resources, storage resources and security level, respectively. The rest numbers represent the available bandwidth resources. We show two possible scenarios in which VNRs \textit{(b)} are embedded in the substrate network. Virtual node \textit{d} may be embedded in substrate node \textit{D} or \textit{H}. Obviously, embedding the substrate node \textit{D} will take up more link bandwidth resources.

In the underlying network, the available node resources can be defined as the current remaining resources, as follows:

\begin{equation}
\begin{aligned}
R_{CPU}(n^s)=CPU(n^s)-\sum_{i=1,n^v \to n^s}^{|VNR|}CPU(n_i^v),
\end{aligned}
\end{equation}
where $R_{CPU}(n^s)$ represents the computing resources available or called remaining resources for the substrate node $n^s$. $\sum_{i=1,n^v \to n^s}^{|VNR|}CPU(n_i^v)$ represents the total amount of CPU resources occupied by virtual nodes embedded on substrate node $n^s$, where the virtual node $n_i^v$ comes from different VNRs.

\begin{equation}
\begin{aligned}
R_{STO}(n^s)=STO(n^s)-\sum_{i=1,n^v \to n^s}^{|VNR|}STO(n_i^v),
\end{aligned}
\end{equation}
where $R_{STO}(n^s)$ represents the remaining storage resources of the substrate node $n^s$. $\sum_{i=1,n^v \to n^s}^{|VNR|}STO(n_i^v)$ represents the amount of storage resources occupied by virtual nodes embedded on $n^s$, where the virtual node $n_i^v$ comes from different VNRs.

Similarly, the current available or remaining bandwidth resources of the substrate link $l^s$ can be expressed as follows.

\begin{equation}
\begin{aligned}
R_{BW}(l^s)=BW(l^s)-\sum_{i=1,l^v \to l^s}^{|VNR|}BW(l_i^v),
\end{aligned}
\end{equation}
where $\sum_{i=1,l^v \to l^s}^{|VNR|}BW(l_i^v)$ represents the bandwidth resource occupied by the virtual links embedded on the physical link $l^s$. The symbol $l^v \to l^s$ represents the collection of all virtual links $l^v$ mapped to the substrate link, where virtual links $l_i^v$ can come from the same VNR.

The restrictive constraints that the VNE problem must follow are as follows.

\begin{equation}
\begin{aligned}
\sum_{i=1}^{|N^V|}num(n_i^v \to n^s)=1,
\end{aligned}
\end{equation}
indicates that a virtual node $n^v$ can only be embedded on one substrate node $n^s$.

\begin{equation}
\begin{aligned}
R_{CPU}(n^s) \geq CPU(n^v).
\end{aligned}
\end{equation}

\begin{equation}
\begin{aligned}
R_{STO}(n^s) \geq STO(n^v).
\end{aligned}
\end{equation}

The above two formulas clarify the computing resource constraints and storage resource constraints of VNE.

\begin{equation}
\begin{aligned}
SL(n^s) \geq SR(n^v).
\end{aligned}
\end{equation}

The above formula clarifies the security constraints of VNE. This security constraint is mainly reflected in the node embedding stage, i.e., the security of substrate node $n^s$ must meet the minimum security level requirements of virtual node $n^v$.

\begin{equation}
\begin{aligned}
\sum_{i=1}^{|N^V|}num(l_i^v \to l^s) \geq 1,
\end{aligned}
\end{equation}
indicates indicates that virtual path segmentation may occur in the link embedding stage.

\begin{equation}
\begin{aligned}
R_{BW}(l^s) \geq BW(l^v),
\end{aligned}
\end{equation}
represents the bandwidth resource constraints of the link embedding stage.

\subsection{Evaluating Indicator}

The ultimate goal of VNE is to receive as many VNRs as possible and to obtain greater benefits of VNE. In order to judge whether a VNE algorithm can achieve the above mentioned excellent performance, some evaluation indexes are usually set up for the algorithm. In the VNE algorithm with computing, storage resources and security constraints, we use three indicators: long-term average revenue, the rate of receiving VNRs and the ratio of long-term revenue consumption to evaluate the algorithm \cite{a4}. For this, we need to make the following definitions.

VNE revenue is the revenue obtained by InPs for providing resource services, defined as:

\begin{equation}
\begin{aligned}
R(G^V)=\sum_{i=1}^{|N^V|}[CPU(n_i^v)+STO(n_i^v)]+\sum_{j=1}^{|L^V|}BW(l_j^v),
\end{aligned}
\end{equation}
where $CPU(n_i^v)+STO(n_i^v)$ represents the computing resources and storage resources occupied by the virtual node $n_i^v$. $BW(l_j^v)$ represents the bandwidth resource occupied by the virtual link $l_i^v$.

The cost consumed by InPs to provide resource services is defined as follows:

\begin{equation}
\begin{aligned}
C(G^V)=\sum_{i=1}^{|N^V|}[CPU(n_i^v)+STO(n_i^v)]+\sum_{j=1}^{|L^V|}\sum_{k=1}^{|L^S|}BW(l_{jk}^{vs}),
\end{aligned}
\end{equation}
where $CPU(n_i^v)+STO(n_i^v)$ represents the computing and storage resources consumed by a virtual node $n_i^v$ after successful mapping. $BW(l_{jk}^{vs})$ represents the cost of bandwidth occupied by $l_{jk}^{vs}$ with link splitting. Virtual link splitting will cause more physical bandwidth consumption, and the cost of VNE will increase.

With the expression methods of revenue and cost, the long-term average revenue of VNE is defined as:

\begin{equation}
\begin{aligned}
AR=\lim_{T \to \infty} \frac{\sum_{t=0}^{T}R(G^V,t)}{T}.
\end{aligned}
\end{equation}

The ratio of long-term revenue consumption is defined as:

\begin{equation}
\begin{aligned}
RC=\lim_{T \to \infty} \frac{\sum_{t=0}^{T}R(G^V,t)}{\sum_{t=0}^{T}C(G^V,t)}.
\end{aligned}
\end{equation}

The VNR acceptance rate is defined as:

\begin{equation}
\begin{aligned}
ACC=\lim_{T \to \infty} \frac{\sum_{t=0}^{T}acc(G^V,t)}{\sum_{t=0}^{T}arrive(G^V,t)},
\end{aligned}
\end{equation}
where $acc(G^V,t)$ represents the number of successfully embedded VNRs. $arrive(G^V,t)$ represents the total number of VNRs that make resource requests to the underlying network.

\section{Algorithm Implementation}\label{part4}

\subsection{Attribute Extraction and Feature Matrix}

Before using RL algorithm, several main elements need to be determined. We use the basic elements of neural network to build a four-layer policy network as the RL agent. The action taken by the agent is the selected VNE strategy. We use the evaluation index of VNE algorithm as reward signal. We first clarify the environment of agent training. In order for the agent to fully learn the complex situation of the underlying network, it is necessary to train the agent in the most realistic network environment. A natural idea is to express the features of the underlying network and normalize these features as an environment for agent training. It is unrealistic to extract all the attributes of the underlying network. If plenty of attributes are extracted, the computational complexity will increase. If too few attributes are extracted, the representation will be incomplete. For this purpose, we extract the following four network features:
\begin{enumerate}
\item Computing resources (CPU): CPU resource is one of the most important resources in network environment. The richer the CPU of the substrate node is, the more virtual nodes it can bear and the greater the probability of being mapped by the virtual node.

\item Storage resources (STO): Storage resources can replace bandwidth resources to a certain extent, rational use of storage resources can release bandwidth pressure and improve the utilization of underlying resources. Storage resource is an important attribute of network node, and bandwidth resource is an important attribute of network link. Replacing bandwidth resource with storage resource can transfer the research of network link to the research of network node, which can effectively reduce the complexity of algorithm.

\item Security level (SL): With the increase of IoT business, network security should be paid attention to. An effective method is to set security level for network nodes to ensure the security of resource scheduling.

\item Average distance of substrate nodes (AVG\_DST): Average distance refers to the average distance between a substrate node and other mapped substrate nodes. When this node is selected for mapping, some bandwidth resources can be saved because it is close to other nodes. The FloydWarshall algorithm is used to calculate the distances to other nodes \cite{10}. The attribute is shown as follows:

\begin{equation}
\begin{aligned}
AVG\_DST(n^s)= \frac {\sum_{i=1,n_i^s \in N^S}^{|N^S|}dis(n^s,n_i^s)}{count+1},
\end{aligned}
\end{equation}
where $\sum_{i=1,n_i^s \in N^S}^{|N^S|}dis(n^s,n_i^s)$ is the distance from the substrate node $n^s$ to the mapped substrate node $n_i^s$.
\end{enumerate}

CPU resources and bandwidth resources should be extracted as important network resources. In order to reduce complexity, we use storage resources instead of bandwidth resources. Setting the security level for network nodes is to ensure the security of resource allocation, thereby ensuring the security of ICPSs. The average distance to other mapped bottom nodes can provide good link embedding options for RL agents. In summary, the above four attributes have broad representative significance. After the node features are extracted, they need to be normalized to the feature vector. For the $k$-th bottom node, its feature vector is as follows:

\begin{equation}
\begin{aligned}
v_{n_k^s}=(CPU(n_k^s),STO(n_k^s),SL(n_k^s),AVG\_DST(n_k^s))^T.
\end{aligned}
\end{equation}

Each substrate node has a corresponding feature vector, which is connected in sequence to form a four-dimensional feature matrix:

\begin{equation}
\begin{aligned}
M_f=[v_{n_1^s},v_{n_2^s},...,v_{n_{|N^S|}^s}].
\end{aligned}
\end{equation}

The feature matrix can be used as the input of the policy network, so that it can ensure that the RL agent learns sufficient network features.

\subsection{Policy Network}

\begin{figure}[!htp]
\centering
\includegraphics[width=1\columnwidth]{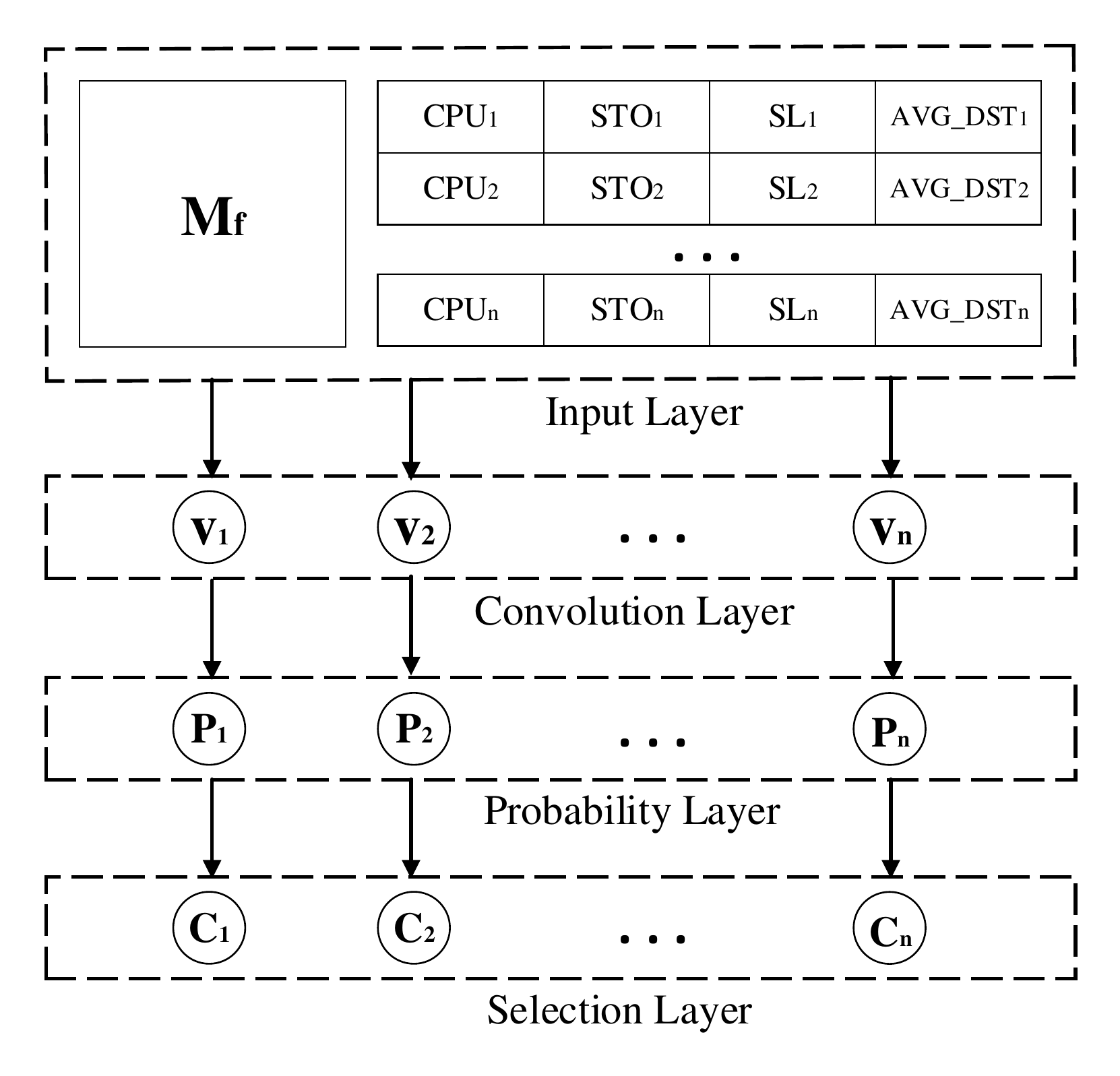}
\caption{Schematic diagram of policy network.}
\label{fig_2}
\end{figure}

Artificial neural network is a common model in ML algorithm. We use its elements to form a simple policy network. Its characteristic is that knowledge or information is stored among all levels of the network. We assume that all VNRs follow the invariant distribution, so we use the feature matrix extraction. The policy network is mainly divided into four layers, as shown in Figure \ref{fig_2}.

The first layer is used to receive the feature matrix extracted from the underlying network. The convolution layer is an important part of the convolution neural network (CNN). Its main function is to evaluate the resource properties of each feature vector in the feature matrix. After comprehensive consideration of the resource amount of the four attributes we extracted, the convolution operation is performed to obtain the available vector form of each node, which is called residual resource vector. The calculation method is as follows:
\begin{equation}
\begin{aligned}
r_i= \omega \cdot v_{n_i} + o,
\end{aligned}
\end{equation}
where $\omega$ is the kernel weight vector, $v_{n_i}$ is the feature vector of the $i$-th substrate node, and $o$ is the offset.

In the probability layer, we mainly use the normalized exponential function, namely softmax function. Its function is to normalize the gradient logarithm of the discrete probability distribution of finite terms. In our network model, the final output is the embedded probability of each underlying node. The probability of the substrate nodes being mapped is positively related to the probability. The generation basis is the residual resource vector of each substrate node. The function of the selection layer is to filter out the substrate nodes that may be due to insufficient resources or security level. These nodes will not be candidate mapping nodes.

\subsection{Training and Testing}

In the training phase, whenever a VNR arrives, the policy network will extract a feature matrix from underlying network. Through a series of operations of the policy network, a set of possible embedded substrate nodes and their probabilities will be generated for each virtual node in the VNR. After embedding all the virtual nodes, breadth first search (BFS) strategy is used to embed the virtual links. In the process of training, there will be a reward problem for agent, so it should choose a suitable standard as the reward signal. For convenience, we use the VNE revenue consumption ratio as the reward signal after the agent takes action. Because this indicator can take into account the revenue and cost of the VNE process at the same time, it can also reflect the utilization of the underlying network resources. The RL agent can decide which action to take according to the magnitude or sign of the reward signal.

We need to set an index to express the speed of training or the magnitude of training range, that is, gradient $g$. The calculation method is as follows:

\begin{equation}
\begin{aligned}
g= \alpha \cdot reward \cdot g_s,
\end{aligned}
\end{equation}
where $\alpha$ is the learning rate. $reward$ stands for reward signal. $g_s$ represents the magnitude of the iterative gradient.

$\alpha$ controls the gradient and the training speed. The  large gradient means that the action conversion range of the agent will be very large and more high-quality solutions may be lost. In addition, too large gradient may lead to unstable network model and incomplete training process. If the gradient is too small, it will lead to extremely slow training process, waste a lot of time and reduce the efficiency of RL. A larger reward signal will lead to a larger gradient. At this time, the agent will be more sensitive and it is easy to take actions to make itself obtain a larger reward.

In order to speed up the parameter updating process and ensure the stability of the process, we choose batch updating method. The specific training process of the VNE algorithm based on RL three-dimensional resource constraints is shown in Algorithm 1.

\begin{algorithm}
  \caption{Training}
  \begin{algorithmic}[1]
     \Require
        {$epochs$, $\alpha$, $train\_set$};
    \Ensure
        {$network\,parameters$};
    \State $parameter\,initialization$;
    \While {$iteration<epochs$}
    \For {$VNRs$}
    \State $counter=0$;
    \For {$virtual\,\,nodes$}
    \State $get\,\,the\,\,feature\,\,matrix\,\,M_f$;
    \State {$whether\,\,resource\,\,constraints\,\,are\,\,met$};
    \State $get\,\,node\,\,probability$;
    \State $computing\,\,the\,\,gradient$;
    \EndFor
    \State $BFS\,\,for\,\,link\,\,mapping$;
    \If {$virtual\,\,links$}
    \State $calculate\,\,reward$;
    \Else
    \State $reset\,\,the\,\,gradient$;
    \EndIf
    \State {$counter++$};
    \EndFor
    \State {$iteration++$};
    \EndWhile
    \State $return\,\,parameters$;
  \end{algorithmic}
\end{algorithm}

The specific testing process of the VNE algorithm based on RL three-dimensional resource constraints is shown in Algorithm 2.

\begin{algorithm}
  \caption{Test process}
  \begin{algorithmic}[1]
     \Require
        {$test\_set$};
    \Ensure
        {$three\,\,evaluation\,\,indicators$};
    \For {$VNRs$}
    \For {$virtual\,\,nodes$}
    \State $get\,the\,feature\,matrix\,M_f$;
    \State $node\,mapping\,based\,on\,probability$;
    \EndFor
    \State $BFS\,for\,link\,mapping$;
    \State {$all\,\,virtual\,\,nodes\,\,and\,\,links\,\,are\,\,embedded$};
    \State $return(SUCCESS)$;
    \EndFor
  \end{algorithmic}
\end{algorithm}

\section{Numerical Results and Analysis}\label{part5}

\subsection{Experimental Environment Setting}

\begin{figure*}[!h]
\centering
\includegraphics[width=1.0\textwidth]{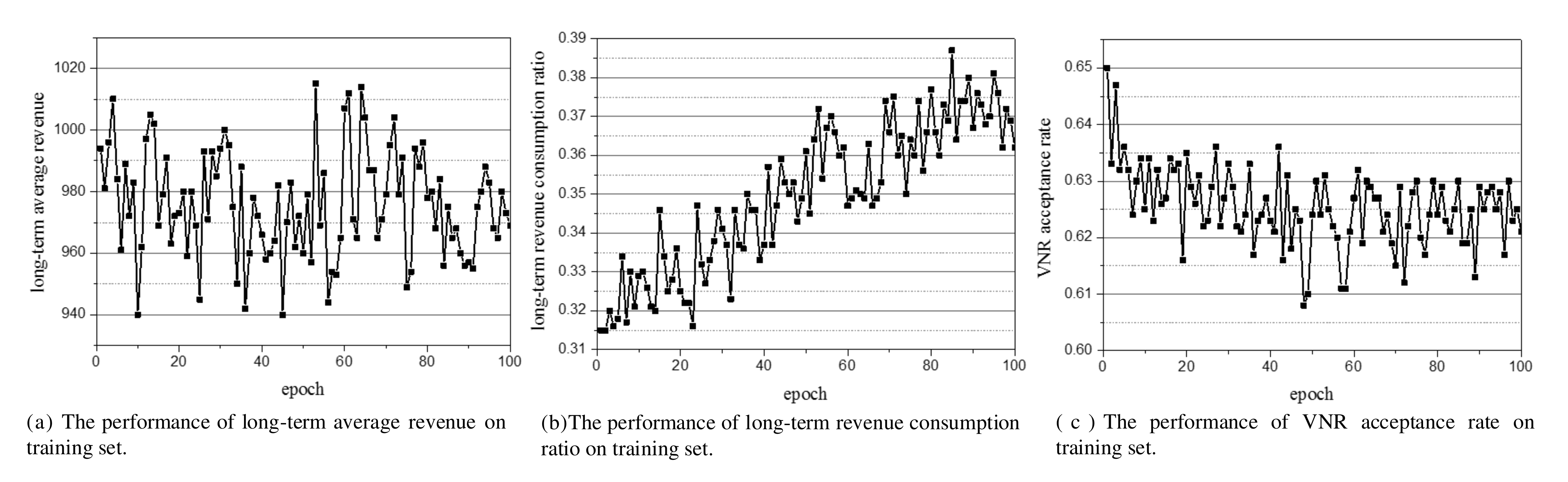}
\caption{Training results.}
\label{fig_3}
\end{figure*}

We have listed the main experimental data settings in Table \ref{tab_1}. The arrival of a VNR simulates the Poisson distribution \cite{12}. Approximately 4 VNRs arrive every 100 units of time. The duration of each VNR follows an exponential distribution, averaging 1,000 units of time overall. We use TensorFlow to construct the policy network and use normal distribution to initialize the parameters of the policy network \cite{14}. Using Anaconda provides a large number of dependency packages for experiments. We set the learning rate to 0.005. Parameters are automatically updated every 100 VNRs received.

\begin{table}
\centering
\caption{Parameter setting of simulation experiment.}
\renewcommand\arraystretch{1.5}
\begin{tabular}{|p{30mm} p{25mm}|}
\hline
Attribute & Value \\
\hline
computing resources & U[50,100]Tflops \\
\hline
storage resources & U[50,100]Mbps \\
\hline
security level & U[0,3] \\
\hline
bandwidth resources &  U[50,100]Mbps \\
\hline
VNRs & 2000 \\
\hline
virtual nodes & U[2,10] \\
\hline
computing requirements & U[0,50]Tflops \\
\hline
storage requirements & U[0,50]Mbps \\
\hline
security requirements & U[0,3] \\
\hline
bandwidth requirements &  U[0,50]Mbps \\
\hline
VNRs for training & 1000 \\
\hline
VNRs for testing & 1000 \\
\hline
\end{tabular}
\label{tab_1}
\end{table}

\subsection{Training Results}

The training results of the above three indicators is shown in Figure \ref{fig_3} respectively. At the beginning of training, since the VNE algorithm solves the NP-hard problem, and the parameters of the policy network are randomly initialized, so it takes some time to adapt to the environment. Therefore, the three indicators fluctuate greatly and the performance is unstable. After a period of training, the performance of the three indicators is somewhat stable. This is because after learning for a while, the agent starts to become familiar with the environment and some actions may lead to a larger reward. But there are also some actions that make the agent get a smaller reward which have less impact on the agent. At the end stage, the agent will explore moves that will give it a bigger reward. From this stage, the effectiveness of RL begins to show. By the end of the training period, the agent has fully adapted to the learning environment and the policy network performance has reached its limit. The training results of the three indexes begin to converge gradually. The agent will get a bigger reward for taking all three measures, so all three indicators start to stabilize. The result is that it fluctuates around a certain value.

The above training process shows that our agent training is effective and the policy network model can play a role. The reason why the agent can achieve good training results is due to the construction of the network environment. The feature matrix we extracted can fully reflect the real resource situation of substrate network, and the agent can perceive the changes of underlying resources in real time to make optimal decisions. Therefore, the training effect is constantly stable and continuously optimized.

\subsection{Test Results}

\begin{figure*}[!h]
\centering
\includegraphics[width=1.0\textwidth]{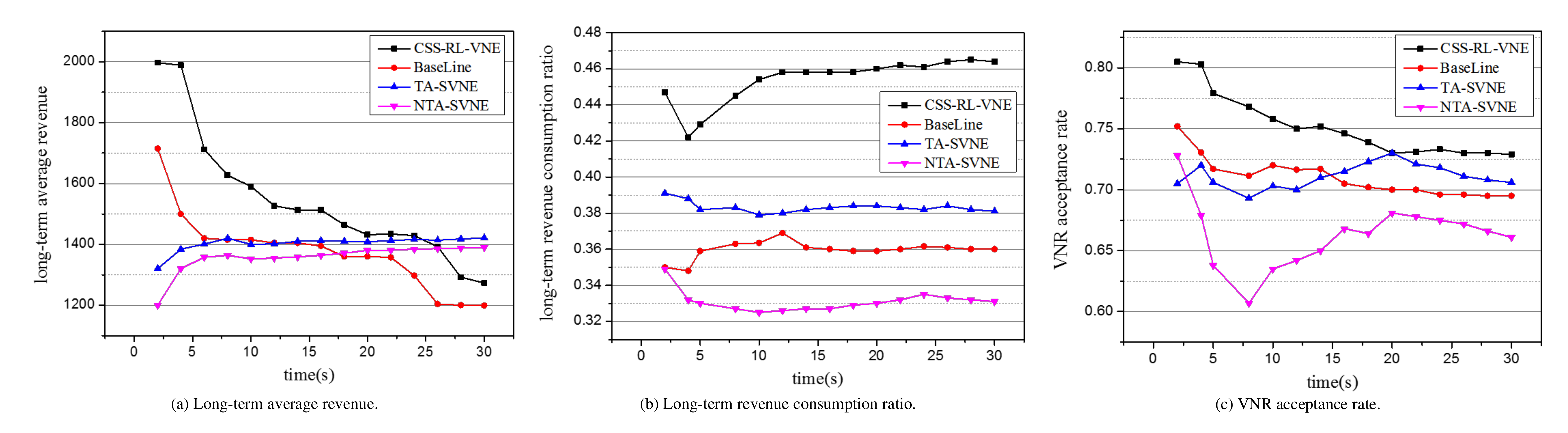}
\caption{Testing results.}
\label{fig_4}
\end{figure*}

In this part, we compare the algorithm for computing, storage and security resource constraints based on RL (CSS-RL-VNE) with three other representative algorithms. In order to compare the fairness, it is necessary to ensure that the initial parameter settings of several algorithms are the same. After the experimental results are obtained, we analyze the results.

The BaseLine algorithm \cite{15} mainly considers the constraints of computing resources and bandwidth resources and sorts the nodes according to the amount of resources. The BaseLine algorithm also uses BFS to complete the process of embedding virtual links. The other two algorithms were proposed in reference \cite{2}, namely TA-SVNE algorithm and NTA-SVNE algorithm. Both of these algorithms are secure VNE algorithms, which focus on the security performance of VNE. In the stage of node mapping, the degree centrality, proximity centrality and resource capability of nodes are used as evaluation indexes. The algorithm uses the approach ideal sorting method to sort the importance of nodes, then maps the virtual nodes according to the sorting results. In the link embedding stage, the algorithm takes the bandwidth resource requirement as the highest priority to embed the virtual link. The method adopted is the k-shortest path first strategy.

Our algorithm is different from them. Firstly, the above three algorithms are all heuristic algorithms and the performance of VNE algorithm is not optimized by means of RL. Secondly, the above three algorithms don't comprehensively consider the impact of computing resources, storage resources and security on VNE. So, our algorithm is innovative.

The experimental results is shown in Figure \ref{fig_4}. Figure \ref{fig_4}-$(a)$ and Figure \ref{fig_4}-$(c)$ both show a downward trend over time. This is because both metrics are related to the quantity status of resources in the underlying network. As the VNRs arrive, the underlying resources are consumed. The long-term revenue consumption ratio is independent of the underlying resource volume. Our algorithm is generally due to the other three algorithms. The reason is that we use efficient RL algorithm. The RL agent can be efficiently trained in a network environment closer to reality and take the action that benefits them the most. So the performance of CSS-RL-VNE algorithm is better than the other algorithms. In summary, the VNE algorithm based on intelligent learning has advantages over the traditional heuristic VNE algorithm.

\section{Summary}\label{part6}

ICPSs and IoT are currently facing severe challenges from resources and security. On the one hand, IoT based on the construction of ICPSs requires a large amount of reasonable network resource support. On the other hand, the rapid development of IoT has also exposed very network security issues. In order to effectively deal with these problems, we proposed a RL algorithm based on the constraints of computing, storage and secure three-dimensional resources from the perspective of VNE. The essence of this algorithm is to make full use of the excellent performance of VNE in resource scheduling. By learning the important attribute characteristics of the underlying nodes, a reasonable node mapping probability is derived, and the node mapping is performed according to this probability. Experimental results show the effectiveness of CSS-RL-VNE algorithm. Accordingly, the effectiveness of the algorithm in supporting the construction of IoT in ICPSs is demonstrated.

\ifCLASSOPTIONcaptionsoff
  \newpage
\fi


\section*{Biographies}

\begin{IEEEbiography}[{\includegraphics[width=1in,height=1.25in,clip,keepaspectratio]{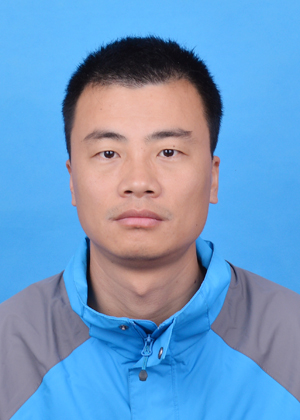}}]{Peiying Zhang}
is currently an Associate Professor with the College of Science and Technology, China University of Petroleum (East China). He is a Ph.D. candidate in Information and Communication Engineering, from the State Key Laboratory of Networking and Switching Technology in Beijing University of Posts and Telecommunications. His research interests include semantic computing, deep learning, network virtualization and future network architecture.
\end{IEEEbiography}

\begin{IEEEbiography}[{\includegraphics[width=1in,height=1.25in,clip,keepaspectratio]{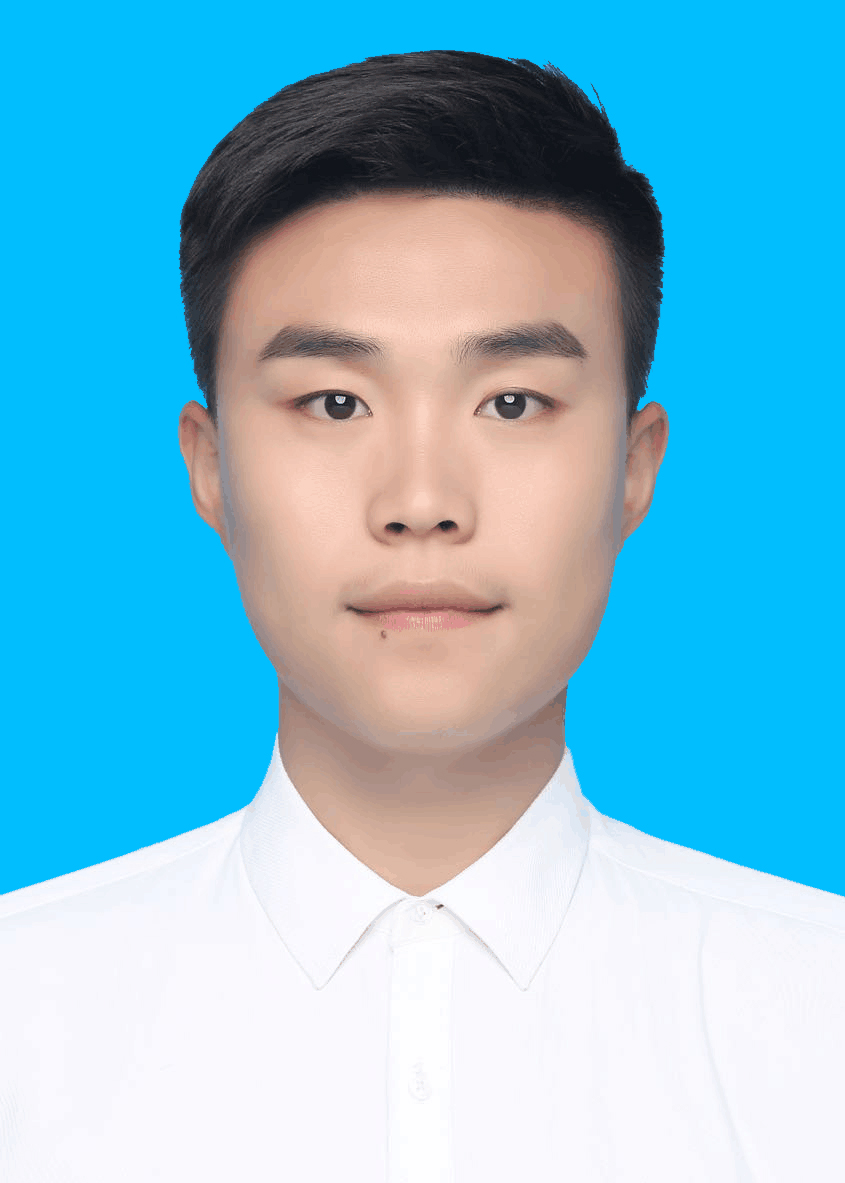}}]{Chao Wang}
is a graduate student in the College of Science and Technology, China university of petroleum (East China). His research interests include network virtualization and network artificial intelligence.
\end{IEEEbiography}

\begin{IEEEbiography}[{\includegraphics[width=1in,height=1.25in,clip,keepaspectratio]{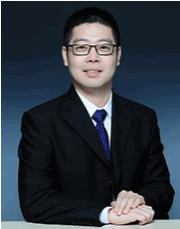}}]{Chunxiao Jiang}
is an Associate Professor in School of Information Science and Technology, Tsinghua University. He received the B.S. degree in information engineering from Beihang University, Beijing in 2008 and the Ph.D. degree in electronic engineering from Tsinghua University, Beijing in 2013, both with the highest honors. His research interests include application of game theory, optimization, and statistical theories to communication, networking, and resource allocation problems, in particular space networks and heterogeneous networks. Dr. Jiang has served as an Editor of IEEE Internet of Things Journal, IEEE Network, IEEE Communications Letters, and a Guest Editor of IEEE Communications Magazine, IEEE Transactions on Network Science and Engineering, and IEEE Transactions on Cognitive Communications and Networking. He has also served as a member of the technical program committee as well as the Symposium Chair for a number of international conferences, including IEEE ICC 2018 Symposium Co-Chair, IWCMC 2018/2019 Symposium Chair, WiMob 2018 Publicity Chair, ICCC 2018 Workshop Co-Chair, and ICC 2017 Workshop Co-Chair. Dr. Jiang is the recipient of the Best Paper Award from IEEE GLOBECOM in 2013, the Best Student Paper Award from IEEE GlobalSIP in 2015, IEEE Communications Society Young Author Best Paper Award in 2017, the Best Paper Award IWCMC in 2017, IEEE ComSoc TC Best Journal Paper Award of the IEEE ComSoc TC on Green Communications \& Computing 2018, IEEE ComSoc TC Best Journal Paper Award of the IEEE ComSoc TC on Communications Systems Integration and Modeling 2018, the Best Paper Award ICC 2019.
\end{IEEEbiography}

\begin{IEEEbiography}[{\includegraphics[width=1in,height=1.25in,clip,keepaspectratio]{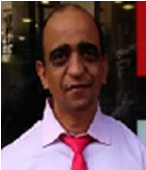}}]{Neeraj Kumar}
received his Ph.D. in CSE from Shri Mata Vaishno Devi University, Katra (Jammu and Kashmir), India in 2009, and was a postdoctoral research fellow in Coventry University, Coventry, UK. He is a full Professor in the Department of Computer Science and Engineering, Thapar Institute of Engineering and Technology (Deemed to be University), Patiala (Pb.), India. He has published more than 300 technical research papers in top-cited journals such as IEEE TKDE, IEEE TIE, IEEE TDSC, IEEE TITS, IEEE TCE, IEEE TII, IEEE TVT, IEEE ITS, IEEE SG, IEEE Netw., IEEE Comm., IEEE WC, IEEE IoTJ, IEEE SJ, Computer Networks, Information sciences, FGCS, JNCA, JPDC and ComCom. He has guided many research scholars leading to Ph.D. and M.E./M.Tech. His research is supported by funding from UGC, DST, CSIR, and TCS. He is an Associate Technical Editor of IEEE Communication Magazine, IEEE Network Magazine. He is an Associate Editor of IJCS, Wiley, JNCA, Elsevier, Elsevier Computer Communications, and Security and Communication, Wiley. He has been a guest editor of various International Journals of repute such as - IEEE Access, IEEE Communication Magazine, IEEE Network Magazine, Computer Networks, Elsevier, Future Generation Computer Systems, Elsevier, Journal of Medical Systems. Springer, Computer and Electrical Engineering, Elsevier, Mobile Information Systems, International Journal of Ad hoc and Ubiquitous Computing, Telecommunication Systems, Springer and Journal of Supercomputing, Springer. He has been a workshop chair at IEEE Globecom 2018 and IEEE ICC 2019 and TPC Chair and member for various International conferences. He is senior member of the IEEE. He has more than 6200 citations to his credit with current h-index of 42. He has won the best papers award from IEEE Systems Journal and ICC 2018, Kansas city in 2018. He has edited more than 10 journals special issues of repute and published four books from CRC, Springer, IET UK, and BPB publications. He is visiting research fellow at Coventry University, Newcastle University, UK.
\end{IEEEbiography}

\begin{IEEEbiography}[{\includegraphics[width=1in,height=1.25in,clip,keepaspectratio]{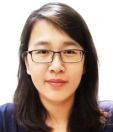}}]{Qinghua Lu}
is a senior research scientist at CSIRO, Australia. She is also a conjoint senior lecturer at University of New South Wales (UNSW). Before she joined CSIRO, she was an associate professor at China University of Petroleum. She formerly worked as a researcher at NICTA (National ICT Australia). She received her Ph.D. from University of New South Wales in 2013. Her research interest includes architecture design of blockchain applications, blockchain as a service, model-driven development of blockchain applications, reliability of cloud computing, and service engineering. She has published more than 70 peer-reviewed academic papers in international journals and conferences. She has served as an editor or reviewer for many journals and as a PC member for a number of international conferences and workshops in the blockchain, cloud computing, big data and software engineering community.
\end{IEEEbiography}

\end{document}